\documentclass[review]{elsarticle}
\usepackage{amsmath,amssymb}
\usepackage{xcolor}

\usepackage{hyperref}

\journal{Journal of \LaTeX\ Templates}

\begin{document}

\begin{frontmatter}
\title{Minimal model for the magnetic phase diagram of CeTi$_{1-x}$Sc$_{x}$Ge, GdFe$_{1-x}$Co$_{x}$Si, and related materials}

\author{William Gabriel Carreras Oropesa$^{1,2}$, Sergio Encina$^1$, Pablo Pedrazzini$^{1,3}$, V\'ictor F. Correa$^{1,3}$, Juli\'an G. Sereni$^{1,3}$, Ver\'onica Vildosola $^{3,4}$, Daniel J. Garc\'{\i}a$^{1,3}$, and Pablo S. Cornaglia$^{1,3}$}

\address{$^1$ Centro At{\'o}mico Bariloche and Instituto Balseiro, CNEA, 8400 Bariloche, Argentina}
\address{$^2$ Instituto de F\'{\i}sica, Universidade de S\~ao Paulo, Caixa Postal 66318, 05314-970 S\~ao Paulo, SP, Brazil}
\address{$^3$ Consejo Nacional de Investigaciones Cient\'{\i}ficas y T\'ecnicas (CONICET), Argentina}
\address{$^4$ Centro At{\'o}mico Constituyentes, CNEA, Buenos Aires, Argentina}

\begin{abstract}
  We present a theoretical analysis of the magnetic phase diagram of CeTi$_{1-x}$Sc$_{x}$Ge and GdFe$_{1-x}$Co$_{x}$Si as a function of the temperature and the Sc and Co concentration $x$, respectively. 
  CeScGe and GdCoSi, as many other RTX (R=rare earth, T=transition metal, X=p-block element) compounds, present a tetragonal crystal structure where bilayers of R are separated by layers of T and X. While GdFeSi and CeTi$_{0.75}$Sc$_{0.25}$Ge are ferromagnetic, CeScGe and GdCoSi order antiferromagnetically with the R 4f magnetic moments on the same bilayer aligned ferromagnetically and magnetic moments in nearest neighbouring bilayers aligned antiferromagnetically. 
The antiferromagnetic transition temperature $T_N$ decreases with decreasing concentration $x$ in both compounds and for low enough values of $x$ the compounds show a ferromagnetic behavior. 
Based on these observations we construct a simplified model Hamiltonian that we solve numerically for the specific heat and the magnetization.
We find a good qualitative agreement between the model and the experimental data. 
Our results show that the main magnetic effect of the Sc $\to$ Ti and Co $\to$ Fe substitution in these compounds is consistent with a change in the sign of the exchange coupling between magnetic moments in neighbouring bilayers. We expect a similar phenomenology for other magnetic RTX compounds with the same type of crystal structure.

\end{abstract}


\end{frontmatter}


\section{Introduction}

Several compounds of the RTX (R=rare earth, T=transition metal, X= p-block element) family crystalize in the CeFeSi-type or CeScSi-type structures \cite{gupta2015review}. These tetragonal structures can be described as stackings of R bilayers separated by layers of T and X. Neutron scattering experiments have found that the magnetically ordered state can generally be described as a stacking of ferromagnetic bilayers coupled ferromagnetically or antiferromanetically between them, depending on the compound \cite{WELTER199954,WELTER199249}. 

Among these compounds, CeScGe crystallizes in the CeScSi-type structure and has attracted considerable attention because of its surprisingly large (for Ce compounds) transition temperature $T_N\simeq 47$ K \cite{sereni2015cdtige}. 
CeTiGe, however, crystalizes in the CeFeSi-type structure and does not present a magnetically ordered state at low temperatures. In recent works, Sereni {\it et al.} studied the evolution of the thermodynamic and transport properties of CeTi$_{1-x}$Sc$_{x}$Ge for samples with $0.25\lesssim x\leq 1$~\cite{sereni2015cdtige,sereni2015exploring,encina2018low}. A continuous reduction of $T_N$ with decreasing Sc content was observed for $x$ down to $\sim 0.5$ and ferromagnetic behavior for lower values of $x$ down to $\sim 0.25$ where there is a change in the crystal structure to CeFeSi-type with no magnetic order.  

The interest in the RTX family has also been fueled by the large magnetocaloric effect (MCE) in the R=Gd compounds. The MCE is generally maximal at temperatures near to the Curie temperature. This makes it attractive for applications to be able to set the transition temperature near the target operation temperature, e.g. by tuning the magnetic exchange couplings. As in the R=Ce compounds, the transition metal T plays an essential role determining the magnetic properties: while GdFeSi and GdCoSi have the same CeFeSi-type structure, GdFeSi is a ferromagnet with $T_C=118$ K and GdCoSi is an antiferromagnet with $T_N=220$ K. The substitution of Co by Fe reduces $T_N$ and leads to a ferromagnetic behavior in GdFe$_{1-x}$Co$_x$Si for $x\lesssim 0.4$ \cite{WLODARCZYK2015273}. Since other RTX compounds like the RFeSi with R=Nd,Sm,Tb also show a change in the nature of the magnetic ground state compared to RCoSi, and all have a CeFeSi-type structure, we would expect for those compounds a qualitatively similar phase diagram to the one observed for GdFe$_{1-x}$Co$_x$Si.

In this work we propose a model to describe the effect of the replacement of the transition metal T by an element in an adjacent column in the periodic table. 
Based on previous results we assume (see Ref.  \cite{vildosola2019magnetic}) that this T replacement produces a local change in the sign of the inter-bilayer exchange coupling. More precisely, that the main effect in the magnetic interactions when a transition metal atom is replaced, is a change in the sign of the magnetic coupling between the two magnetic moments which are closer to the transition metal atom and in different bilayers.
\section{Model}
To analyze the main effects of the transition metal replacement we construct a simplified magnetic model that takes into account the layered structure of the CeScSi-type and CeFeSi-type crystals.  
The nature of the local magnetic moments can vary widely from one rare earth compound to the other. In CeScGe the transition temperature is high enough ($T_N\sim 47$ K) that the first excited crystal split doublet of the Ce 4f orbital cannot be ignored for a detailed description \cite{ritter2016tetragonal}. In the Gd compounds the crystal field is generally very small \cite{betancourth2015evidence,betancourth2019magnetostriction}, and an isotropic spin $7/2$ describes accurately the physics \cite{facio2015why}. Note, however, that a simple de Gennes scaling is nicely followed by the transition temperature for several of the RTX compounds \cite{WELTER199954}, which signals the possibility of a common description.

\begin{figure}[h]
	\begin{center}
  \includegraphics[width=0.4\textwidth]{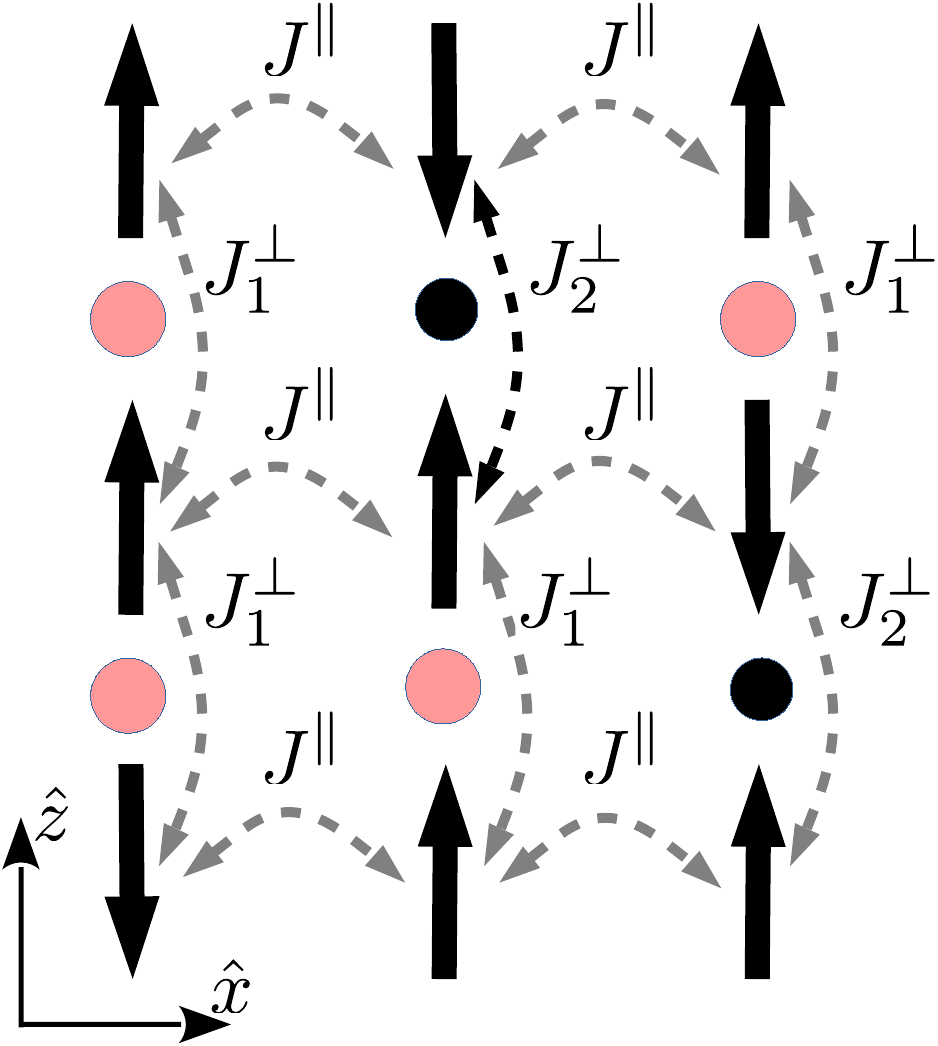}
	\end{center}
	\caption{Schematic representation of the model used in the simulations for a $0<x<1$ situation. The Ising spins form a cubic lattice with ferromagnetic nearest-neighbour couplings $J^\parallel$ in the $\hat{x}-\hat{y}$ plane. The magnetic couplings in the $\hat{z}$-axis ($J_1^\perp$ or $J_2^\perp$) depend on the type of transition metal atom (represented by filled circles) between the magnetic moments.}
  \label{fig:model}
\end{figure}

In constructing the simplified model we do not attempt to reproduce the precise spin arrangement nor the complexity of the rare earth magnetic moments for each compound, but to account for the main parameters driving the magnetic characteristics.
We consider a cubic array of Ising magnetic moments with a first neighbour ferromagnetic coupling $J^\parallel$ inside the $\hat{x}-\hat{y}$ planes and a coupling $J^\perp$ for nearest neighbours in the $\hat{z}$ direction. For a description of the GdCoSi and CeScGe compounds, $J^\perp$ is chosen antiferromagnetic which leads to an A-type antiferromagnetic ground state as observed experimentally. To describe the ferromagnetic GdFeSi a ferromagnetic $J^\perp$ needs to be considered.
In agreement with the double exchange mechanism for the magnetic couplings across the TX layer described in the Appendix, we assume that the most relevant effect of the replacement of a T atom in a TX layer is a change in the coupling between the nearest neighbour R 4f magnetic moments to the T ion across the TX layer. For the Co $\to$ Fe and the Sc $\to$ Ti replacements (which change the parity of the T 3d level occupancy) it can even be associated with a sign change in the interplane coupling $J^\perp$ (see also Ref. \cite{vildosola2019magnetic}).
For a compound where a proportion $x$ of the T atoms have been replaced, we expect an equal proportion of interplane couplings to change. We assume the replaced T atoms to be randomly distributed throughout  the sample which leads to a uniform random distribution in the location of the modified couplings (see Fig. \ref{fig:model}). 

\subsection{Ising model}

For a cubic lattice with magnetic moments at positions $(i,j,k)$ with $i$ ,$j$, and $k$ natural numbers in the range $[1,L]$ the model Hamiltonian reads 
\begin{align}
	H=&J^\parallel \sum_{\langle (i,j), (i^\prime,j^\prime)\rangle} \sum_k  S(i,j,k)S(i^\prime,j^\prime,k)\\
	+&\sum_i \sum_j \sum_{ k} J^\perp(i,j,k) S(i,j,k)S(i,j,k+1)\\
	+& \frac{\Delta}{2} \sum_{i,j,k} S(i,j,k)
	\label{eq:ising}
\end{align}
where  $S(i,j,k)=\{-1,1\}$ represents an Ising spin, $\Delta$ is a Zeeman energy splitting due to an external magnetic field, $\langle,\rangle$ indicates nearest neighbors and $J^\perp(i,j,k)$ can take two different values: $J_1^\perp$ with probability $x$ and $J_2^\perp$ with probability $1-x$. For fixed $J^\perp(i,j,k)$ the model can be analyzed as a function of the temperature using Monte Carlo simulations. 

\section{Results}
We first perform a mean field analysis averaging over disorder realizations of the couplings. This leads to a uniform $J^\perp_{eff}(x)= x J_1^\perp+(1-x)J_2^\perp$ coupling in the $\hat{z}$ direction. The effective interplane coupling $J^\perp_{eff}(x)$ changes sign for $x=x_c\equiv\frac{1}{1-\frac{J_1^\perp}{J_2^\perp}}$, with $0\leq x_c\leq 1$ provided $J_1^\perp$ and $J_2^\perp$ have opposite sign. At $x_c$ the ground state changes from FM to A-type AFM. In what follows we take $J^\perp_2=4 J^\parallel<0$ \footnote{We obtained qualitatively similar results for different values of the ratio between $J^\perp_2$ and $J^\parallel$.} and $J_1^\perp=-3J_2^\perp$ which leads to $x_c=0.25$.

The magnetic transition temperature in the mean field approximation is given by
\begin{equation}
  T_{crit}(x)=\frac{2}{k_B}\left[2 |J^\parallel| + |J^\perp_{eff}(x)|\right]
  \label{eq:tcmf}
\end{equation}
which for $x\leq x_c$ is a Curie temperature to a ferromagnetic ground state and for  $x>x_c$ corresponds to a N\'eel temperature.

\begin{figure}[h]
	\begin{center}
  \includegraphics[width=0.5\textwidth]{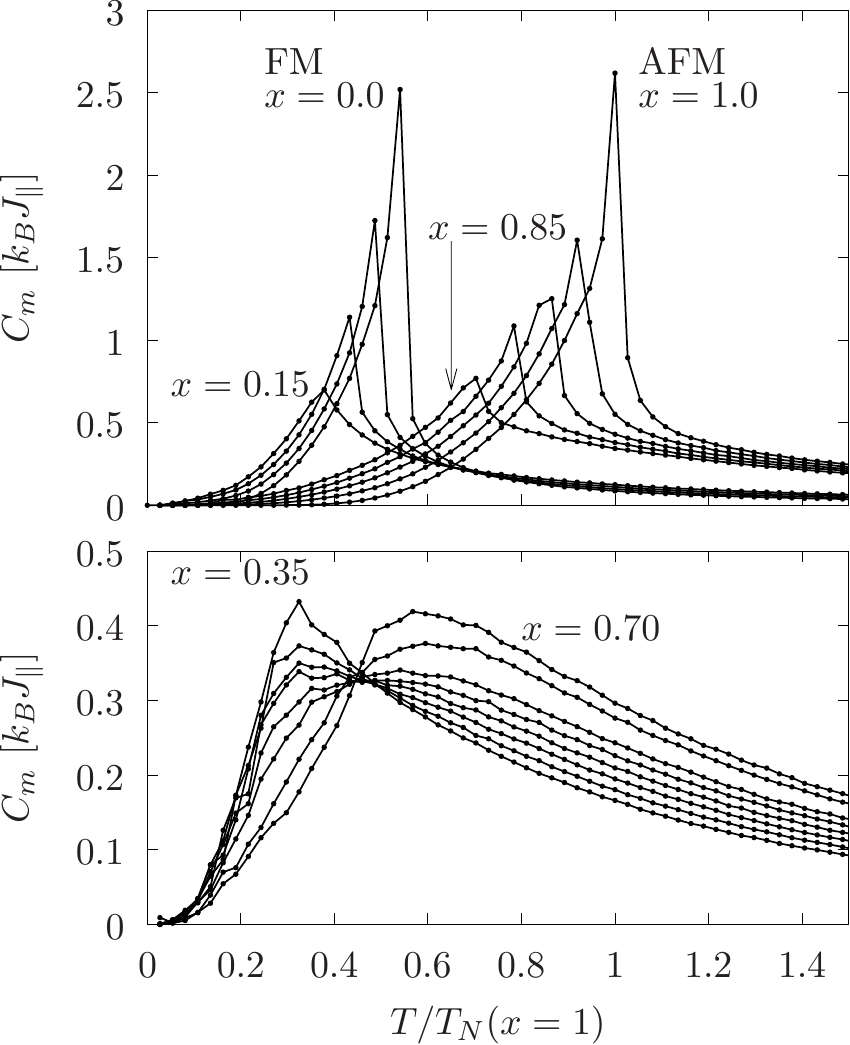}
  \caption{Specific heat $C_m$ of the disordered Ising model as a function of the temperature for different proportions $x$ of modified couplings in the $\hat{z}$ direction. Upper panel: For two sets of values of $x$: $\{0,\,0.05,\,0.1,\,0.15\}$ and $\{0.85,\,0.9,\,0.95,\,1\}$ a sharp peak can be observed in $C_m$ which can be associated with a ferromagnetic and an antiferromagnetic transition, respectively. Lower panel: for $0.35\lesssim x\lesssim 0.7$ the peak in $C_m$ is rounded and no clear signature of a sharp transition is observed.}
	\label{fig:cv}
	\end{center}
\end{figure}
To describe the effect of disorder in a more realistic way we perform Monte Carlo simulations as a function of the temperature and external magnetic field for systems with sizes up to $N=32\times32\times32$ magnetic moments in a cubic lattice with periodic boundary conditions, and averaged the results over three realizations of the disorder (the specific heat and the magnetization show little dependence on the disorder realization for the system sizes considered).
The results for the specific heat as a function of the temperature and different values of the concentration $x\sim 1$  and to $x\sim 0$ are presented in the upper panel of Fig. \ref{fig:cv}. 
The sharp peak in $C_m$ at the transition temperature obtained for $x=1$ and $x=0$ shifts to lower temperatures as $x$ departs from these values. 
A broadening of the peaks is also observed, in agreement with the reported specific heat experimental data for CeTi$_{1-x}$Sc$_x$Ge and $x\sim 1$. 
In the intermediate regime of values of $x$, a broad peak is observed in the specific heat (see lower panel of Fig. \ref{fig:cv}).

\begin{figure}[h]
	\begin{center}
 \includegraphics[width=0.7\textwidth]{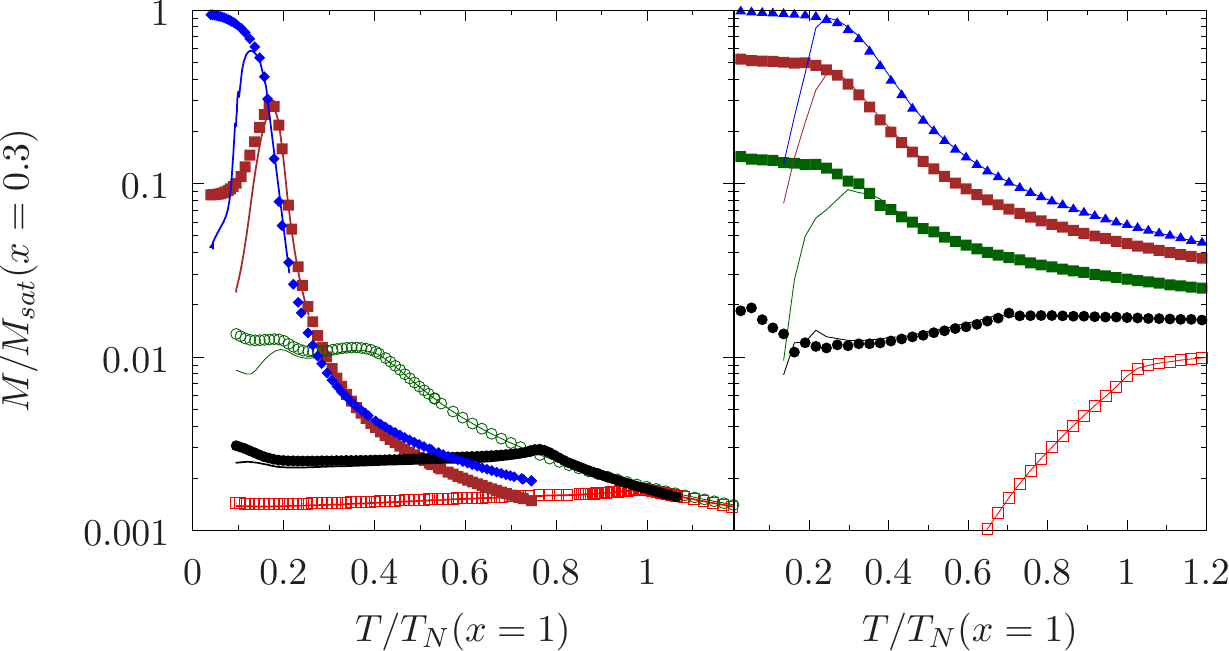}
 \caption{Magnetization as a function of the temperature for the transition metal concentration values $x=\{0.3,\,0.4,\,0.6,\,0.8,\,1.0\}$ from top to bottom at low temperatures using a field cooling (symbols) and zero field cooling (solid lines) schemes. Left panel: experimental data for CeTi$_{1-x}$Sc$_x$Si and $H=100$Oe. Right panel: Monte Carlo data for a Zeeman splitting $\Delta=0.02 J^\parallel$. }
 \label{fig:mag}
	\end{center}
\end{figure}
We also calculated the magnetization as a function of the temperature using a field cooling (FC) and a zero-field cooling protocol. The results of the Monte Carlo simulation and the experimental results for CeTi$_{1-x}$Sc$_x$Si are presented in Fig. \ref{fig:mag}. The numerical calculations and the experiment present a similar qualitative behavior for all values of $x$, except for $x=0.4$ where the FC experimental results present a peak in the magnetization. The observed reduction of the magnetization at low temperatures may be due to a dipolar interaction between ferromagnetic clusters which is not included in the model. For values of $x\sim 0$ ($x\sim 1$), the peak in the specific heat is concomitant with an increase (decrease) in the magnetization which signals a ferromagnetic (antiferromagnetic) transition. In the intermediate range of values of $x$, although there is a clear change in the behavior of the magnetization at temperatures where a peak in the specific heat is observed, it is not apparent from these quantities what the nature of the magnetic ground state is and whether there is a sharp transition.
\begin{figure}[h]
	\begin{center}
  \includegraphics[width=0.7\textwidth]{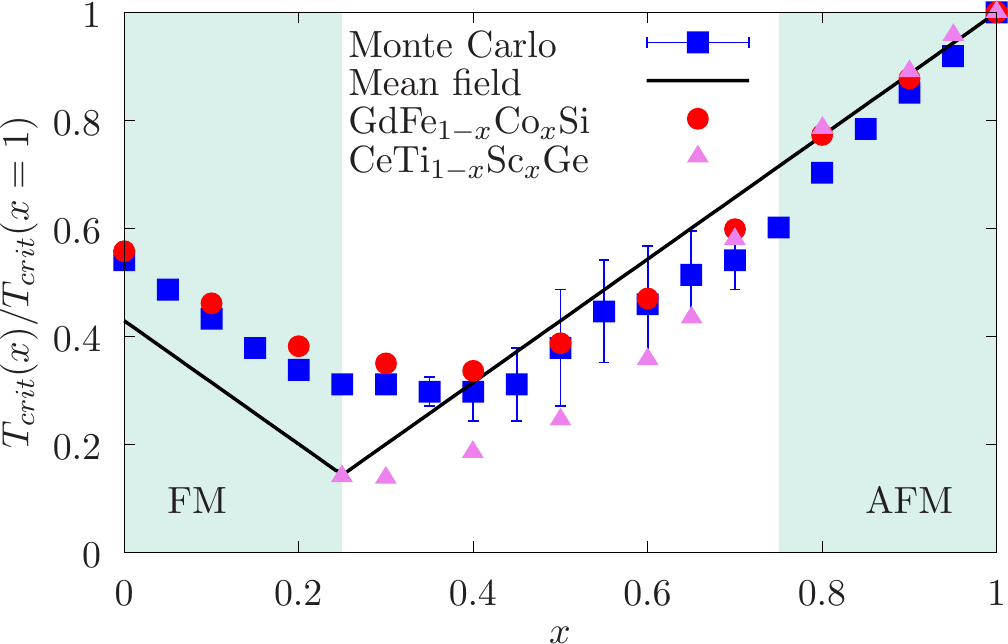}
	\end{center}
	\caption{Experimental transition temperatures for CeTi$_{1-x}$Sc$_x$Ge (from Ref. \cite{WLODARCZYK2015273}) and GdFe$_{1-x}$Co$_x$Si (from Ref. \cite{sereni2015cdtige}) as a function of $x$. The solid line is the mean-field transition temperature for disorder-averaged couplings, to an FM ground state for $0\leq x\leq 0.25$ and an AFM ground state for $0.25< x\leq 1$. The Monte Carlo values are the average of the temperatures for the maximum values of $C_m$ and $C_m/T$, the error bars are estimated from the difference of the values obtained by the two criteria.}
  \label{fig:pd}
\end{figure}

Finally, we use the peaks in the specific heat $C_m$ and in $C_m/T$ as criteria to determine the transition (or crossover) temperature $T_{crit}$. In the intermediate regime of values of $x$, the two criteria do not coincide and we use this difference to estimate the ``error'' in the determination of $T_{crit}$. The resulting $T_{crit}$, which is the average of the value obtained using the two criteria, is presented in Fig. \ref{fig:pd} as a function of $x$, together with the experimental values for CeTi$_{1-x}$Sc$_{x}$Ge and GdFe$_{1-x}$Co$_{x}$Si. In spite of the simplicity of the model, a good agreement is obtained between the Monte Carlo and the experimental results. The mean field results (using an effective disorder-averaged interplane coupling) deviate significantly from the numerical results in the range of intermediate values of $x$, but provide a qualitative correct picture with a linear behavior in $x$, for $x\sim 0$ and $x\sim 1$.

\section{Conclusions}
We presented a minimal model to describe the magnetic properties of CeTi$_{1-x}$Sc$_{x}$Ge and GdFe$_{1-x}$Co$_{x}$Si, which provides a qualitative description of the magnetic specific heat and the magnetization as a function of the temperature. Using Monte Carlo simulations we obtained a magnetic phase diagram which shows a good agreement with those observed experimentally.
The proposed model captures the most relevant effect of the Sc $\to$ Ti and Co $\to$ Fe replacements in these materials, which is a change in the sign of the exchange coupling between R 4f magnetic moments in neighbouring bilayers. Although a material specific model is probably needed to describe the detailed physics of other compounds with the same type of crystal structure but different rare earth ions with their corresponding multiplet structures, we expect the present model to serve as a base to include this type of features.

The similarity of the model presented to the Anderson model (see e.g. Ref. \cite{nishimori2001statistical}), which presents a spin glass phase, suggests that this type of physics could be expected for CeTi$_{1-x}$Sc$_{x}$Ge and GdFe$_{1-x}$Co$_{x}$Si  and $x\sim 0.5$. 

This work opens the possibility to analyze the properties of the low temperature state in the intermediate range of values of the transition metal concentration $x$ using a simple model.

We acknowledge insightful discussions with A. Kolton and financial support from PICT 2016-0204.

\section*{Appendix: simplified model for the interplane R-R exchange coupling}
We present here a simplified local model for the exchange coupling of nearest-neighbouring R magnetic moments in different bilayers. 
As discussed for R=Gd compounds in Ref. \cite{vildosola2019magnetic}, the most relevant effect in the magnetic couplings when the transition metal is replaced (Ti $\to$ Sc or Fe $\to$ Co), is a significant modification, that can even lead to a sign change, of the exchange couplings between the magnetic moments of nearest-neighbour Gd$^{3+}$ ions in different bilayers. A DFT analysis indicates an important overlap of the T 3d, Gd  5d and X p wavefunctions, which leads to an indirect magnetic exchange mechanism via delocalized 5d rare earth electrons \cite{nikitin1996itinerant} (see also Ref. \cite{facio2015why}). 
Additionally, the R $4f$ electrons couple with the R $5d$ conduction electrons with a magnetic exchange coupling $J_{fd}$ and the almost empty R $5d$ orbitals have a small hybridization with the partially occupied transition metal $d$ orbitals. 

With these ingredients we construct simplified model to describe qualitatively the behavior of interlayer exchange coupling. We consider two R ions (1 \& 2) separated by a transition metal ion and consider a single level with energy $E_{\alpha,\sigma}$ for the each of the R $5d$ orbitals $\alpha=1,2$. The R 5d orbitals are hybridized with a single effective level with energy $E_d$ that models the transition metal $d$ orbitals.

While the Si or Ge p orbitals contribute to the conduction electron bands, and to the R-R exchange couplings, we focus here in the role of the TM and do not include the p orbitals in the model.
The energy of an electron in the 5d orbital depends on the relative orientation of its spin $\sigma$ w.r.t the R 4f magnetic moment: $E_{d\sigma}=E_{d}\pm \sigma \delta$, where the - (+) sign corresponds to parallel (antiparallel) configurations.  The R-R coupling is estimated as $K_\perp \sim (E_{AP}-E_{P})/2J^2$, where $E_P$ is the electronic energy when the 4f spins of the two R are parallel, and $E_{AP}$ the corresponding to the antiparallel configuration.
The model Hamiltonian is
\begin{equation}
	H_{eff} = \sum_{\alpha,\sigma} E_{\alpha\sigma}c_{\alpha,\sigma}^\dagger c_{\alpha,\sigma}^{} + E_{d}\sum_\sigma d_{\sigma}^\dagger d_{\sigma} + t\sum_{\alpha,\sigma} (c_{\alpha,\sigma}^\dagger d_\sigma + h.c.)
    \label{toymodel}
\end{equation}
where $c_{\alpha,\sigma}^\dagger$ ($d_{\sigma}^\dagger$) creates an electron with spin projection $\sigma=\pm$ along the z-axis on the R $5d$ (T $3d$) effective orbital.
The model can be readily diagonalized and for, a single electron occupancy, to lowest order in $t$ and $\delta$:
\begin{equation}
	K_\perp\sim \frac{8\delta^2 t^4}{J^2 (E_c-E_d)^5} <0
	\label{K2}
\end{equation}
where we have assumed that $t$ is a small parameter. This corresponds to a ferromagnetic interaction. 

For an occupancy of two electrons there is a qualitative change. The two electrons are antiparallel to satisfy Pauli exclusion principle and the minimal energy for the conduction band electrons is obtained when the R 4f magnetic moments are antiparallel.
This leads to an effective antiferromagnetic exchange coupling between the R magnetic moments in adjacent planes.
\begin{equation}
  K_\perp\sim \frac{|\delta| t^2}{J^2 (E_c-E_d)^2}>0.
	\label{K2}
\end{equation}

\section*{References}

\bibliography{references}

\end{document}